# Happiness and the productivity of software engineers

- Daniel Graziotin, Institute of Software Technology, University of Stuttgart, Germany.
- Fabian Fagerholm, Blekinge Institute of Technology, Sweden and Department of Computer Science, University of Helsinki, Finland.

## Abstract

Software companies nowadays often follow the idea of flourishing happiness among developers. Perks, playground rooms, free breakfast, remote office options, sports facilities near the companies… There are several ways to make software developers happy. The rationale is that of a return on investment: happy developers are supposedly more productive and, hopefully, also retained.

But is it the case that *happy software engineers = more productive software engineers*? [1] Moreover, are perks the way to go to make developers happy? Are developers happy at all? These questions are important to ask both from the perspective of productivity and from the perspective of sustainable software development and well-being in the workplace.

This chapter provides an overview of our studies on the happiness of software developers. You will learn why it is important to make software developers happy, how happy they really are, what makes them unhappy, and what is expected for their productivity while developing software.

## Key Ideas

- Science says the industry should strive for happy developers.
- The overall happiness of software developers is slightly positive. Yet, many are still unhappy.
- The causes of unhappiness among software engineers are numerous and complex.
- Happiness and unhappiness bring a plethora of benefits and detriments to software development processes, people, and products.

## Why the industry should strive for happy developers

We could think that happiness is a personal issue that individual developers are responsible for on their own time. In this line of thinking, software companies should focus on maximizing the output they get from each developer. However, to get productive output from a human, we must first invest. As humans, software developers' productivity depends on their skills and knowledge – but to access those, we need to create favorable conditions that allow the human potential to be realized. It follows that companies should be interested in the general well-being of their software developers. Furthermore, we believe we should simply strive to create better working environments, teams, processes, and therefore products.

# What is happiness and how do we measure it?

This is a very deep question that ancient and modern philosophers have aimed to answer in more than one book. However, present-day research does give us concrete insight into happiness and ways to measure it. We define happiness (as many others do) as a sequence of experiential episodes. Being happy corresponds to frequent positive experiences, which lead to experiencing positive emotions. Being unhappy corresponds to the reverse: frequent negative experience leading to negative emotions. Happiness is the difference or balance between positive and negative experiences. This balance is sometimes called *affect balance*.

The Scale of Positive and Negative Experience (SPANE, [8]) is a recent but valid and reliable way to assess the affect balance (happiness) of individuals. Respondents are asked to report on their affect, expressed with adjectives that individuals recognize as describing emotions or moods, from the past four weeks. This provides a balance between the sampling adequacy of affect and the accuracy of human memory to recall experiences and reduce ambiguity. The combination of the scoring of the various items yields an affect balance (SPANE-B) score, which ranges from –24 (extremely unhappy) to +24 (extremely happy), where 0 is to be considered a neutral score of happiness.

# Scientific grounds of happy and productive developers

While it is intuitive that happiness is beneficial for productivity and well-being, these ideas are also supported by scientific research. We have previously shown that happy developers solve problems better [1], that there is a relationship between affect and how developers assess their own productivity [2], and that software developers themselves are calling for research in this area [5]. We have also presented a theory which provides an explanation of how affect impacts programming performance [3]: events trigger affects in programmers. These affects might earn importance and priority to a developer's cognitive system, and we call them attractors. Together with affects, attractors drive or disturb programmers' focus, which impacts their performance. On a larger scale, our studies show that affect is an important component of performance in software teams and organizations [11]. Affect is linked to group identity – the feeling of belonging to the group – affecting cohesion and social atmosphere, which in turn are key factors for team performance and retention of team members.

We will now consider four important and ambitious questions:

1. How happy are software developers overall?
2. What makes them (un)happy?
3. What happens when they are (un)happy?
4. Are happy developers more productive?

Answering these questions is very challenging. We spent a year designing a comprehensive study [4,6] to address them. We needed data from as many software developers as possible. We also needed as much diversity as possible in terms of age, gender, geographical location, working status, and other background factors. We designed and piloted a questionnaire in such a way that the results could be generalizable (with a certain error tolerance) to the entire population of software developers. Our questionnaire had demographic questions, SPANE, and open-ended questions asking about developers' feelings of happiness and unhappiness when developing software. We asked them to describe a concrete recent software development experience, what could have caused them to experience their feelings in that situation, and if their software development was influenced by these feelings in any way, and how.

We obtained 1318 complete and valid responses to all our questions.

## How happy are software developers?

In the following figure, you can see how happy our 1318 participants were.

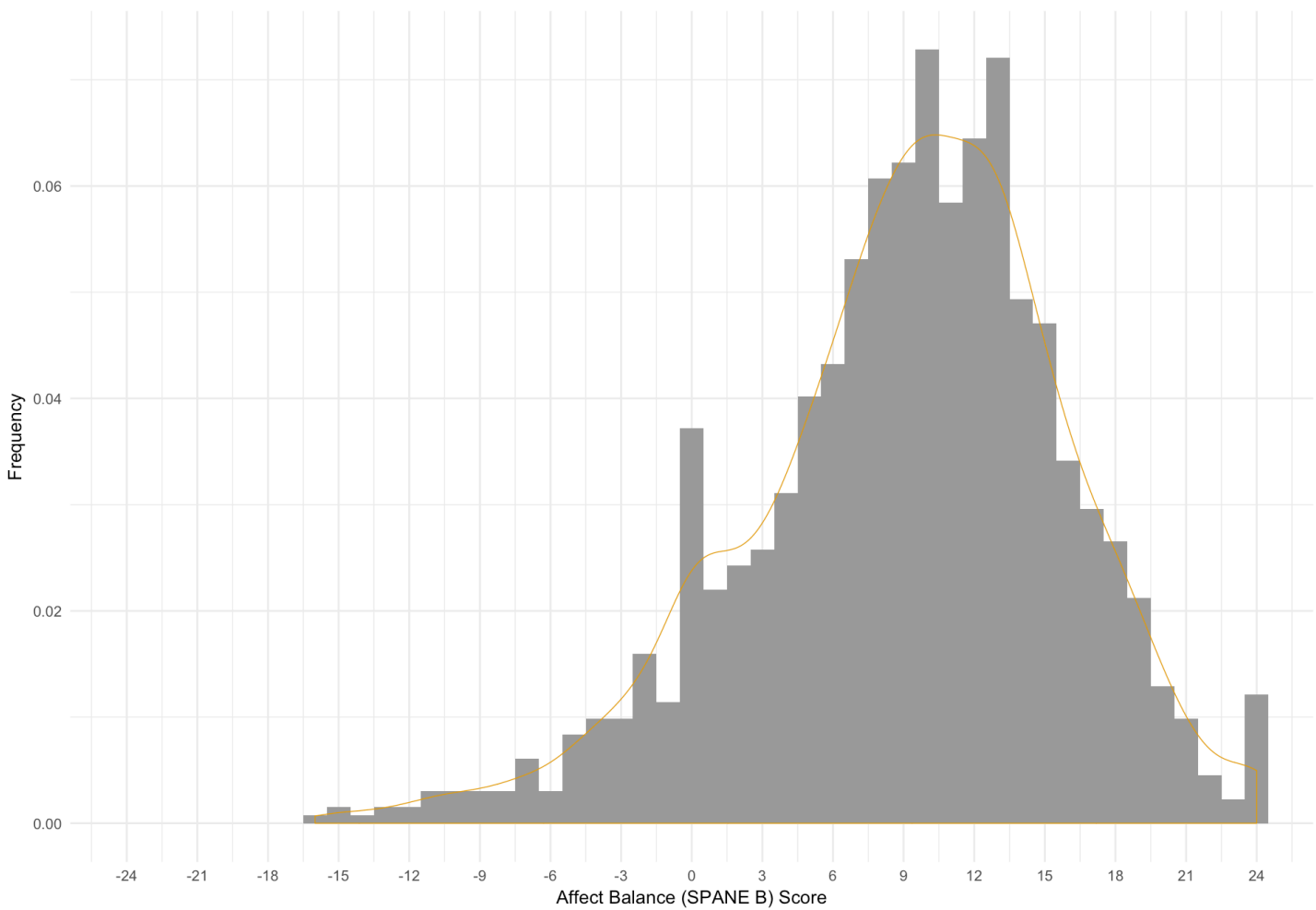

Our participants had a SPANE-B (see *What is happiness and how do we measure it?*) average score of 9.05, and we estimated the true mean happiness score of software developers to be between 8.69 and 9.43 with a 95% confidence interval. In other words, most software developers are moderately happy.

We compared our results with similar studies (Italian workers, US college students, Singapore university students, Chinese employees, South African students, and Japanese college students). All results from other studies report a mean SPANE-B score higher than 0, but lower than in our study. Software developers are indeed a *slightly happy* group – and they are happier than what we would expect based on knowledge about various other groups of the human population. This is good news, indeed, but there is room for improvement, nonetheless. Some developers have a negative SPANE-B score, and there were many examples in the open responses about episodes of unhappiness that could be avoided.

## What makes developers unhappy?

Our analysis of the responses of our 1318 participants uncovered 219 causes of unhappiness, which were mentioned 1843 times in the responses [4]. As the main focus of this chapter is not about how to make developers happy or not unhappy, we present here a brief summary of the results.

The causes of unhappiness that are controllable by managers and team leaders, seem to have an incident rate of 4 times more than those coming from the private sphere. We also expected the majority of the causes coming from human related considerations. However, most of them came from technical factors from the artifact (software product, tests, requirements and design document, architecture, etc.) and the process. This highlights the importance of strategic architecture and workforce coordination.

Being stuck in problem solving and time pressure are the two most frequent causes of unhappiness, which corroborates the importance of recent research that attempts to understand them. We recognize that it is in software development's nature to be basically problem-solving under deadlines, but the issue does not lie in this. We cannot avoid problem solving in software development. However, developers *feel bad* when they are stuck and under pressure, and several detrimental consequences do happen (see the rest of this chapter). This is where we researchers (and then *you*, managers) should intervene to reduce the detrimental effects of these issues. Psychological grit could be an important characteristic to train among software developers. Another could be how to switch your mindset to get unstuck.

The third most frequent cause of unhappiness is to work with bad code and, more specifically, with bad code practices. Developers are unhappy when they produce bad code, but they suffer tremendously when they meet bad code that could be avoided in first place. Often, our participants stated, this is an issue that is mandated by management for saving time and effort. The same could be said of third persons (could be colleagues, as well as team leaders or customers) that make developers feel inadequate with their work; forced repetitive mundane tasks; and imposed limitations on development. Much of this can be avoided by task rotation, by taking better decisions, and by actually listening to developers. Several top causes are related to the perception of inadequacy of the self and others, which encourages recent research activities on intervening on the affect of developers [3].

Finally, we see that factors related to information needs in terms of software quality and software construction are strong contributors to unhappiness among developers. More research is needed on producing tools and methods that make communication and knowledge management of software teams easier to store, retrieve, and comprehend.

## What happens when developers are (un)happy?

When presenting early results of our study, we produced the following representation of the consequences of unhappiness of software developers.

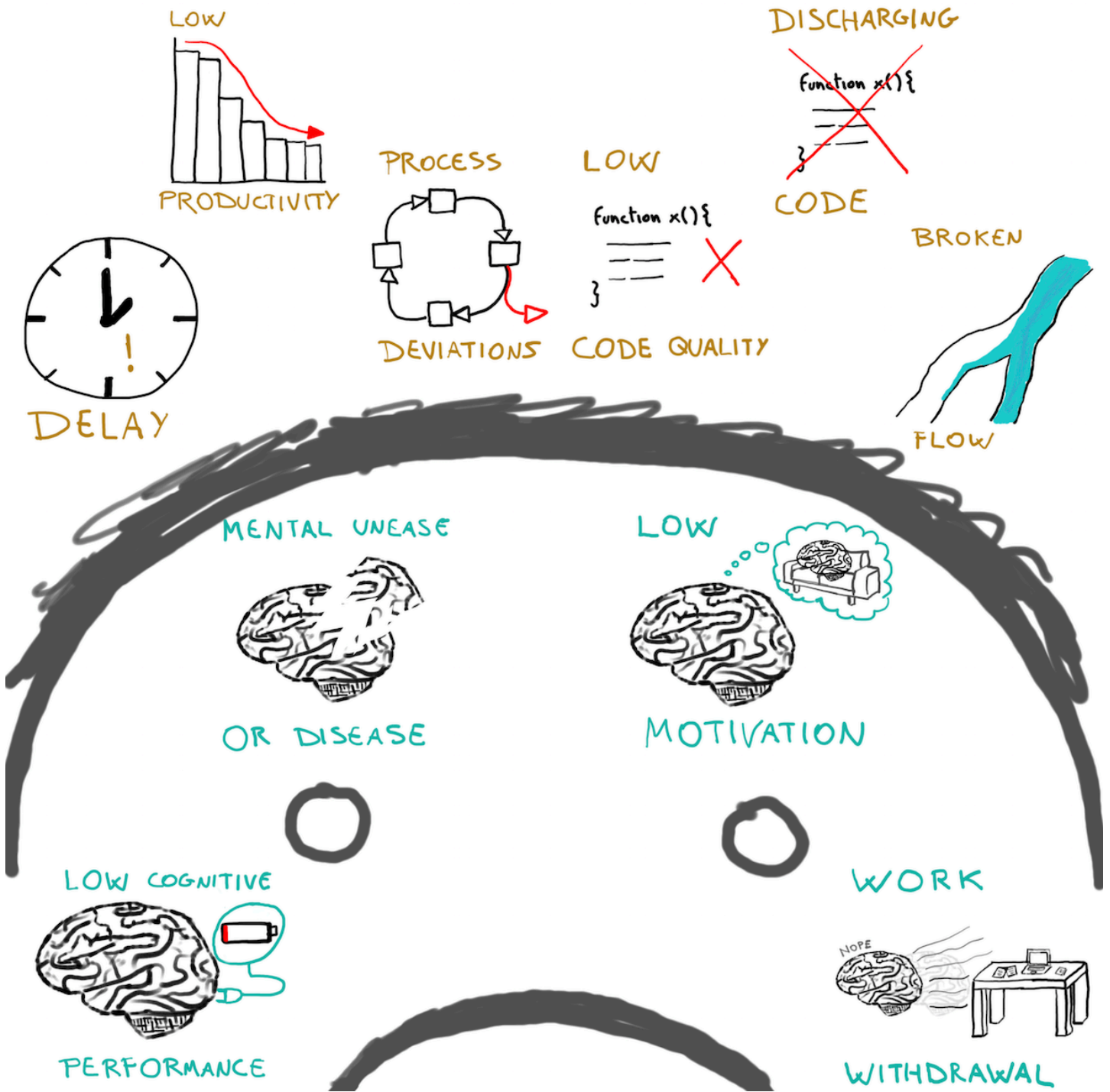

*Consequences of unhappiness while developing software. Available as CC-BY from Graziotin, Daniel; Fagerholm, Fabian; Wang, Xiaofeng; Abrahamsson, Pekka (2017): Slides for the consequences of unhappiness while developing software. figshare. https://doi.org/10.6084/m9.figshare.4869038.v3*

As you can see, developers reported a variety of productivity-related consequences – and some even explicitly reported experiencing lower productivity. Other consequences include delays, process deviations, and low code quality. These external effects are direct impacts on productivity and performance. Internal consequences, such as low motivation and cognitive performance indirectly affect productivity as well. We coded the open-ended questions and found dozens of causes and consequences of happiness and unhappiness while developing software [4, 6].

For the purposes of this chapter, it is worth going into more detail on the consequences of happiness and unhappiness, because several of them are productivity-related and productivity was the most populated category of consequences. We are reporting them in an order that favors narrative, not by frequency of occurrence.

## Cognitive performance

We found that being happy or unhappy influences several factors related to cognitive performance, that is, how we efficiently process information in our brain.
Happiness and unhappiness influence how we can focus while coding, as put by one participant: "[…] the negative feelings lead to not thinking things through as clearly as I would have if the feeling of frustration was not present". The opposite holds true: "My software development is influenced because I can be more focused on my tasks and trying to solve one problem over another". As the focus can be higher when happy (or lower when unhappy), a natural consequence is that problem-solving abilities are influenced: "I mean, I can write codes and analyze problems quickly and with lesser or no unnecessary errors when I'm not thinking of any negative thoughts".
Being happy while developing software brings higher learning abilities: "It made me want to pursue a masters in computer science and learn interesting and clever ideas to solve problems". However, being unhappy causes much mental fatigue, and participants reported to "Getting frustrated and sloppy".

## Flow

Participants mentioned how being unhappy caused breaks in their flow. Flow is a state of intense attention and concentration resulting from task-related skill and challenge being in balance. Unhappiness causes interruptions in developers' flow, resulting in adverse effects on the process. As put by a participant, "things like that [of unhappiness] often cause long delays, or cause one getting out of the flow, making it difficult to pick up the work again where one has left off. ". When happy, developers can enter a state of sustained flow. They feel full of energy and with strong focus. In such a state, they are "unaware of time passing". They can "continue to code without anymore errors for the rest of the day", and "just knock out lines of code all day", with "dancing fingers, my code is like a rainbow".

## Motivation and withdrawal

Motivation was often mentioned by our participants. They were clear in stating that unhappiness leads to low motivation for developing software: "[the unhappiness] has left me feeling very stupid and as a result I have no leadership skills, no desire to participate and feel like I'm being forced to code to live as a kind of punishment". The participants also stated that increased motivation occurred as they were happy.

Unhappiness and happiness are causes of work withdrawal and work engagement, respectively. Work withdrawal is a very destructive consequence of unhappiness, and it emerged often among the responses. Work withdrawal is a family of behaviors that is defined as employees' attempts to remove themselves, either temporarily or permanently, from daily work tasks.

We found varying degrees of work withdrawal, ranging from switching to another task: "[…] you spend like 2 hours investigating on Google for a similar issue and how it was resolved, you find nothing, desperation kicks in. It clouds your mind and need to do other things to clear it", to considering quitting developing software, "I really start to doubt myself and question whether I'm fit to be a software developer in the first place", or even quitting the job.

High work engagement and perseverance, on the other hand, were reported to occur when respondents were happy. This means, for example, pushing forward with a task: "I think I was more motivated to work harder the next few hours". This is slightly different from motivation, which is more about the energy directed to acting towards a goal. Work engagement is committing to the act of moving towards a goal.

## Happiness, unhappiness, and how they relate to productivity of developers

Finally, **participants directly mentioned how unhappiness hinders their productivity**. We grouped all responses related to performance and productivity losses. The codes within this category were ranging from very simple and clear "productivity drops", "[negative experience] definitely makes me work slower" to more articulated "[unhappiness] made it harder or impossible to come up with solutions or with good solutions". Unhappiness also causes delays in executing process activities: "In both cases [negative experiences] the emotional toll on me caused delays to the project".

Of course, participants reported that **happiness leads to high productivity**: "When I have this [happy] feeling I can just code for hours and hours", "I felt that my productivity grew while I was happy", "The better my mood, the more productive I am".

More details on that by one participant: "I become productive, focused and enjoy what I'm doing without wasting hours looking here and there in the code to know how things are hooked up together".

An interesting aspect is that, when happy, developers tend to take on undesired tasks: "I think that when I'm in this happy state I am more productive. The happier I am the more likely I'll be able to accomplish tasks that I've been avoiding".

On the other hand, unhappy developers could be so unproductive that they become destructive. We found some instances of participants who destroyed the task-related codebase, e.g., "I deleted the code that I was writing because I was a bit angry", up to deleting entire projects: "I have deleted entire projects to start over with code that didn't seem to be going in a wrong direction".

Another intriguing aspect is about long-term considerations of being happy: "I find that when I feel [happy], I'm actually more productive going into the next task and I make better choices in general for the maintenance of the code long-term. […] I'm more likely to comment code thoroughly".

## Are happy developers more productive?

Alright, but *are happy developers* **really** *more productive*? Whenever science attempts to show if a factor *X* causes an outcome *Y*, researchers design *controlled experiments*. Controlled experiments attempt to keep every possible factor constant (*A*, *B*, *C*, …) except for the factors (*X*) that should cause a change to the outcome *Y*. Whenever this control is not possible, we call these studies *quasi experiments*.

Here is the issue with research on happiness. It is challenging to control the happiness (or the mood, the emotions) of people. One of the reasons is that a perfect controlled experiment would need to be quite unethical to make the unhappy control group… truly unhappy. The effects of asking participants to remember sad events, or showing depressing photographs, is negligible. Still, we could setup two quasi experiments to observe some correlations.

One of these studies [1] has received considerable media attention. We tested a hypothesis regarding a difference of intellectual (cognitive-driven) performance in terms of analytical (logical, mathematical) problem solving of software engineers according to how happy they were.
We also wanted to perform a study where all the tools and measurements came from psychology research and were validated.
So, we designed a quasi-experiment in a laboratory, where 42 BSc and MSc students of computer science had their happiness measured and then conducted a task resembling algorithmic design. For measuring happiness we opted for SPANE (explained above).

The analytic task was similar to algorithm design and execution. We decided to administer the Tower of London test (also known as Shallice test) to our participants. The Tower of London test resembles the Tower of Hanoi game. The test comprises two boards with stacks and several colored beads. There are usually three stacks per board and each stack can accommodate only a limited number of beads. The first board presents predefined stacked beads. The participants received the second board, which has the same beads as the first board but stacked in a different configuration. The participants have to recreate the configuration of the first board by unstacking one bead at a time and moving it to another stack. The [Psychology Experiment Building Language (PEBL)](#) is an open source language and a suite of neuropsychology tests [13], [14]. The Tower of London test is among them.

PEBL was able to collect the measures that let us calculate a score for the analytic performance. We compared the scores obtained in both tasks with the happiness of developers. The results showed that the happiest software developers outperformed the other developers in terms of analytic performance.
We estimated the performance increase to be of about 6%. The performance increase was not negligible, and we confirmed it by measuring Cohen's *d* statistic. Cohen's *d* is a number usually ranging from 0 to 2, which represents the magnitude of the effect size of a difference of means. Our Cohen's *d* for the difference between the two groups mean was 0.91, a large effect – given that we did not obtain extreme cases of happiness and unhappiness. The margins could even be higher than that.

In another study [2], we did something more esoteric. We aimed to continue using psychology theory and measurement instruments for understanding the linkage between the real-time affect (let's say happiness) raised by a software development task and the productivity related to the task itself. Eight software developers (4 students, 4 from software companies) worked on their real-world, software project. The task length was 90 minutes (as it is about the typical length for a programming task). Each 10 minutes, the developers filled a questionnaire formed by the Self-Assessment Manikin (SAM) and an item for self-assessing the productivity.

SAM is a scale for assessing an emotional state or reaction. SAM is peculiar because it is a validated way to measure the affect raised by a stimulus (like an object, or a situation) and it is picture-based (no words). SAM is simply three rows of puppets with different face expressions and body language. Therefore, it is very quick for a participant to fill SAM especially if implemented on a tablet device (only 3 touches). We analyzed how developers felt during the task and how they self-assessed themselves in terms of productivity. Self-assessment is not a very objective way of measuring productivity, but it has been demonstrated that individuals are actually good at self-assessing themselves if they are observed alone [15]. The results have shown that high pleasure with the programming task and the sensation of having adequate skills are positively correlated with the productivity. This correlation holds over time, and real-time. We also found that there are strong variations of affect in 90 minutes of time. Happy software developers are indeed more productive.

## Potential impacts of happiness on other outcomes

Happiness influences so many things besides productivity, most of which are still related to development performance. Here we list three of them.

Unhappiness causes glitches in communication and a disorganized process: "Miscommunication and disorganization made it very difficult to meet deadlines". But happy developers can also mean more collaborative team members, leading to increased collaboration. Often, we saw a repeating pattern of willingness to share knowledge ("I'm very curious and i like to teach people what i learned") and to join an effort to solve a problem ("we never hold back on putting our brains together to tackle a difficult problem or plan a new feature") even when not related to the task at hand or the current responsibilities ("I was more willing to help them with a problem they were having at work.").

Being happy or unhappy does not only influence the productivity of the code writing process but also the quality of the resulting code. Participants reported that "eventually [due to negative experiences], code quality cannot be assured. So this will make my code messy and more bug can be found in it", but also mentioned making the code less performant, or "As a result my code becomes sloppier". Sometimes, being unhappy results in discharging quality practices, for example: "[…] so I cannot follow the standard design pattern", as a way to cope with the negative experiences.
Yet, being happy improves the quality of code. A participant told a small story about their work: "I was building an interface to make two applications talk. It was an exciting challenge and my happy and positive feelings made me go above and beyond to not only make it functional but I made the UX nice too. I wanted the whole package to look polished and not just functional".
When happy, developers tend to make less mistakes, see solutions to problems more easily, and make new connections to improve the quality of the code. A participant told us: "When I'm in a good mood and I feel somehow positive, the codes I write seems to be very neat and clean. I mean, I can write codes and analyze problems quickly and with lesser or no unnecessary errors". As a result, the code is cleaner, more readable, better commented and tested, and with less errors and bugs.

The last factor we would like to report is mostly related to unhappiness, and it is quite an important one. It is about mental unease and mental disorder. We created this category to collect those consequences that threaten mental health. Participants reported that unhappiness while developing software is a cause of anxiety: "These kinds of situations make me feel panicky"; stress: "[…] only reason of my failure due of burnout"; self-doubt: "If I feel particularly lost on a certain task, I may sometimes begin to question my overall ability to be a good programmer"; and sadness and feeling depressed, e.g., "feels like a black fog of depression surrounds you and the project" or "I get depressed". In addition, we found mentions of feelings of being judged, frustration, and lack of confidence in one's ability.

# What does the future hold?

In 1971, Gerald Weinberg's book "The psychology of programming" [12] drew attention to the fact that software development is a human endeavor, and the humans – the developers – doing it are individuals with feelings. To this day, much remains in understanding the human factor in software development. Many avoidable mistakes are still made because of failures to consider the strengths and limitations of humans. Software development productivity is still often managed as if it were about delivering code on an assembly line. On the other hand, many companies do understand the importance of happy developers, invest into their well-being, and consider it to be worthwhile. It is important that we learn more about the relationship between well-being and software development performance. Rigorous research, and educating practitioners on the research results, are keys to improve the field.

To this end, we envision a future where both of these problems have begun to be addressed. We want to explore the relationship between developers' happiness and development outcomes such as productivity, code quality, and impact of the software systems developed. This would open possibilities for new ways to design and validate software engineering methods so that they are maximally and sustainably usable for real developers and organizations. Simultaneously, we believe that advances in research will foster the next generation of software developers through improved education. Besides sharp technical skills, we would like to give future software developers an understanding of the social and psychological factors that influence their own work.

As we have shown, the link between happiness and productivity in software development is real. It is possible to quantify the happiness of software developers, and there are distinct patterns in the causes and consequences of their happiness. What if we could include happiness as a factor in software development productivity management? In the future, an increasing number of people will work with digital products and services and perform tasks that are, in effect, software development. It would be worth investing in their happiness.

# Further reading

In this book chapter, we reported on several studies on the happiness of software engineers. Some of these studies [1, 2, 3, 5, 11] were self-contained and independent. Other studies [4, 6] are part of an ongoing project that we described in the first part of Section *Scientific grounds of happy and*

*productive developers*.

At the time of writing of this book chapter, we still have to uncover all categories, including those about what makes developers happy. We invite the readers to inspect our open science repository [10], where we add new papers and results as we uncover them. The repository contains the entire taxonomy of what makes developers unhappy.

---

1. In our studies, we consider a software developer to be "a person concerned with any aspect of the software construction process (such as research, analysis, design, programming, testing, or management activities), for any purpose including work, study, hobby, or passion." [4, page 326]. We also interchange the terms "software developer" and "software engineer" so that we do not repeat ourselves too many times. ↵